\documentclass[article,12pt,3p]{elsarticle}
\biboptions{numbers,sort&compress} 
\usepackage[
singlelinecheck=false 
]{caption}
\usepackage{enumitem}
\usepackage{bm}
\usepackage{setspace}
\usepackage{makecell}
\usepackage{multirow}
\usepackage[dvipsnames,usenames]{xcolor}
\usepackage{amsmath}
\usepackage{amssymb}

\journal{\mbox{ \ }}
\begin{document}
	\makeatletter
	\def\ps@pprintTitle{%
		\let\@oddhead\@empty
		\let\@evenhead\@empty
		\def\@oddfoot{\reset@font\hfil}%
		\let\@evenfoot\@oddfoot}
	\makeatother
	
	\renewcommand{\arraystretch}{1.3}

	\begin{frontmatter}
		
		\title{One-dimensional $AB$ random sequential adsorption with one deposition per site}
			
		
		\author[label1]{Charles S. do Amaral\corref{cor1}}
		\address[label1]{Departamento de Matem\'atica - Centro Federal de Educação Tecnológica de Minas Gerais.}
		\ead{charlesmat@cefetmg.br}
		\cortext[cor1]{Corresponding author.}
  
\author[label2]{Diogo C. dos Santos}
\address[label2]{Instituto de Matemática - Universidade Federal de Alagoas.}
		
		\date{\today}
		
		\begin{abstract}

 We investigate a modified version of the $AB$ random sequential adsorption model. Specifically, this model involves the deposition of two distinct types of particles onto a lattice, with the constraint that different types cannot occupy neighboring sites. By restricting the deposition attempts to only one per site, we derive an analytical expression for the average densities of particles of types $A$ and $B$, at all time instances, for all deposition probabilities of each particle type. 
		\end{abstract}
		
	\end{frontmatter}

\section{Introduction}

Random sequential adsorption (RSA) is a mathematical model that describes the behavior of particles as they adsorb onto a surface. It has been widely used to study various physical, chemical, and biological phenomena, such as the deposition of thin films \cite{ramsden_thin_film1, sorrell_thin_film2}, the formation of biological membranes \cite{krugmann_membrane1,Lu_membrane2,carton_membrane3} and the adsorption of molecules onto surfaces \cite{molecules1_Li,molecules2_Erban,molecules3_ciesla1,molecules4_ciesla2}. In the RSA, particles are sequentially added to a surface according to certain rules, and the probability of a particle adsorbing onto the surface is determined by the spatial arrangement of previously adsorbed particles. A detailed overview of the RSA are given by Penrose and Sudbury \cite{penrose}, and Evans \cite{evans}.

Throughout this work, we consider a version of the \textit{Two-species antagonistic random sequential adsorption model}, studied in \cite{martins}, on the one-dimensional lattice. In this model, particles $A$ and $B$ are deposited randomly on a lattice with the constraint that different particles cannot occupy nearest-neighbor sites. Deposition attempts at a site occur until it is occupied or deposition is no longer possible. Therefore several deposition attempts may occur at each site.

In the model studied by us, we consider only one deposition attempt per site, more precisely, we can define the model in the following steps.

\begin{itemize}
	\item[(1)] For each site $s \in \mathbb{Z}$ assign a number $t_s$ randomly and uniformly chosen from the interval $[0,1]$.
 
 	\item[(2)] Fix $\alpha$ and $\beta$ in $[0,1]$ so that $\alpha+\beta=1$. For each site $s$, choose a particle type, denoted by $Y_s$, to attempt deposition on it. The chosen particle will be $A$ or $B$ with probabilities $\alpha$ and $\beta$, respectively.

	\item[(3)] Consider the continuous time process where at time $t=0$ all sites are vacant. Each site $s$ will be occupied in time $t_s$ by a particle of type $Y_s$, chosen at random in Step~2, provided that the neighboring sites $s-1$ and $s+1$ are not already occupied with particles of the opposite type. When $t=1$ we have already tried to occupy all the sites, and the model reaches a frozen state (\textit{jammed state}).
\end{itemize}

\noindent Note that if we denote by $X$ an empty site, then a configuration of the type $AXA$ can occur in the jammed state. Already in the model with several deposition attempts per site, this setting will inevitably change to $AAA$ at some moment.

We will denote by $\rho_{A}(t; \alpha)$ and $\rho_{B}(t; \alpha)$ the average density of sites occupied by $A$ and $B$, respectively, at time $t$. The average density of empty sites at time  $t$ will be denoted by $\rho_{X}(t; \alpha)$. In this paper, we calculate $\rho_{A}(t; \alpha)$ analytically for all $\alpha,t\in[0,1]$. We demonstrate that
\begin{align}\label{principal}
    \rho_{A}(t; \alpha) =  \dfrac{1}{4\theta} \left[2\theta+ \left(\theta^2 - 1\right)2\gamma t + \left(\theta^2+1\right)\sinh(2\gamma t) - 2\theta\cosh(2\gamma t) \right],
\end{align}
where $\gamma=\sqrt{\alpha \beta}$ and $\theta=\sqrt{\dfrac{\alpha}{1-\alpha}}$.

Observe that, as $\rho_{A}(t; \alpha)=\rho_{B}(t; 1-\alpha)$ and $\rho_{A}(t; \alpha)+\rho_{B}(t; \alpha)+\rho_{X}(t; \alpha)=1$, for all $t, \alpha \in [0,1]$, our result gives the average density of each type of particle for all $t$ and $\alpha$ values. In the case that $\alpha=\beta$, it follows from \eqref{principal} that  
 \begin{equation*}
\rho_A(t;\alpha)=\rho_B(t;\beta)=\tfrac{1}{2}(1-e^{-t}).
 \end{equation*}

 We employ an approach similar to that used in \cite{gerin} and \cite{amaral_santos}, where the average density of open sites was calculated at any time $t$ for the \textit{parking process} and the \textit{RSA model with nearest neighbor exclusion}, respectively. In the later model, the authors also obtained the exact expression of the pair correlation function.

In the model with more than one deposition per site, the authors in \cite{martins} obtained results for the average densities in the jammed state on the square lattice when considering low values of $\beta$. In addition, other results related to the percolation of particles of certain types are presented.


The subsequent sections of this paper are organized as follows. Section 2 provides a description of the main events used to determine the probability of a site being occupied by a particle of type $A$. The calculation of the probabilities associated with these events is presented in Section 3. In Section 4, we derive an expression for $\rho_{A}(t;\alpha)$. The behavior of this function is then analyzed in Section 5. Finally, Section 6 summarizes the conclusions drawn from our findings.

\vskip 1cm

\section{Main events}

In this section, we will define mutually exclusive events $\mathcal{G}_{j,k}^{oo}, \mathcal{G}_{j,k}^{eo}, \mathcal{G}_{j,k }^{oe}$ and $\mathcal{G}_{j,k}^{ee}$, for $j,k \geq 0$, such that $\rho_{A}(t; \alpha)$ will be equal to the probability of the union of all them.
We will denote by $\mathbb{P}$ the measure of probability associated with the described model and by $\mathcal{O}_{t}^{s}$ the event in which the site $s$ is occupied by a particle of type $A$ at time $t$. Since $\mathbb{P}(\mathcal{O}_{t}^{s})$ takes on the same value regardless of the site $s$, it is held that $\rho_{A}(t; \alpha)=\mathbb{P}(\mathcal{O}_{t}^{0})$.

\begin{figure}[h!]
	\begin{center}
		\includegraphics[width=16.2cm, height=9.5cm]{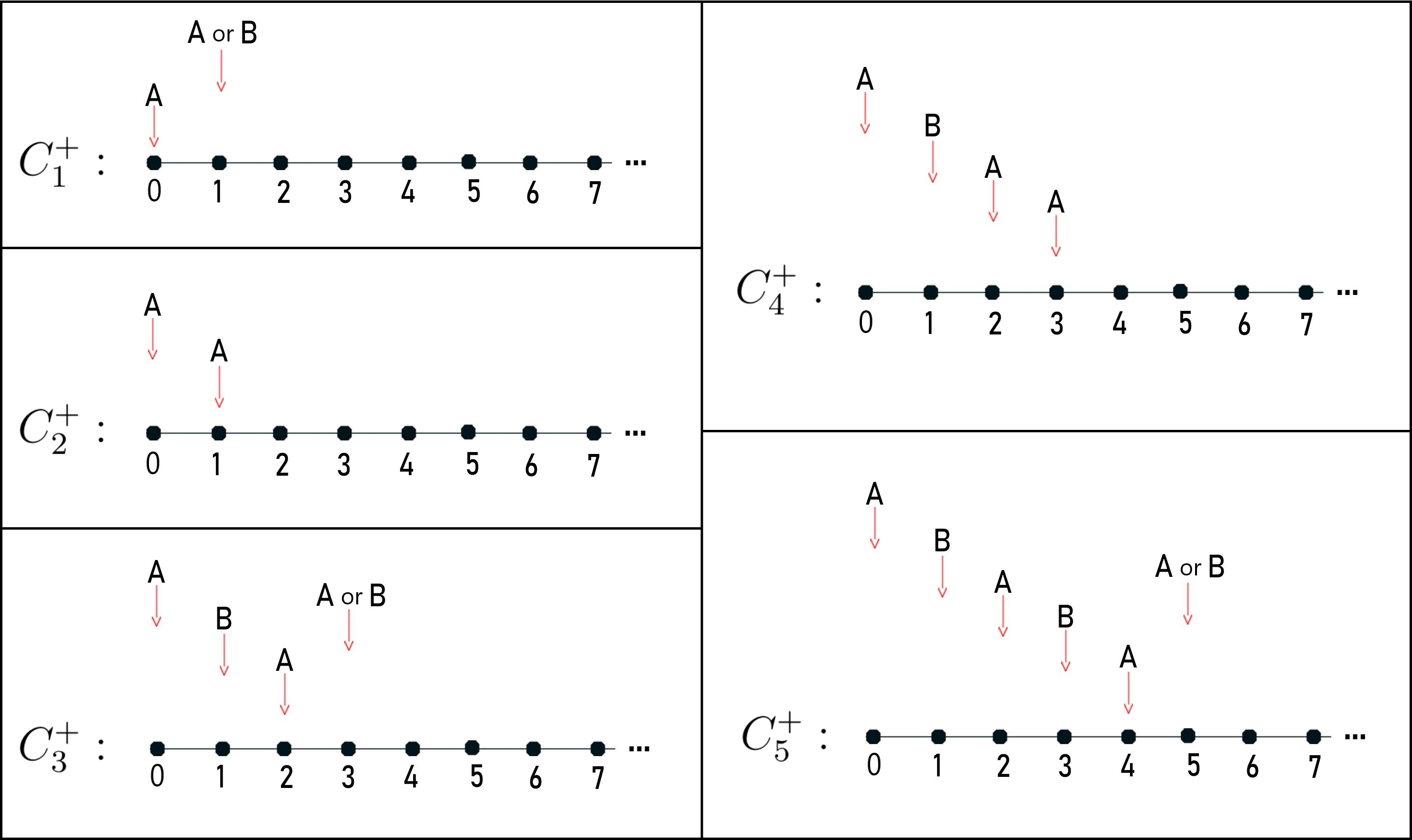}
		\caption{Occurrence of the events $C_i^{+}$, for $i=1,2,3,4,5$. The letters represent the type of particle that we will try to deposit in that site.
The deposition attempt order follows the order of greatest proximity to the graph. In sites where it is not indicated which particle can be deposited, there may be an attempt to deposit any type of particle and at any instant of time $t$.}
		\label{fig_deposition}
	\end{center}
\end{figure}

We start by considering the model restricted to sites $s\in\{0,1,...\}$ and define mutually exclusive events $C_k^{+}$, $k \geq 1$,  such that the occurrence of any one of them implies the occurrence of $\mathcal{O}_{t}^{0}$. To this end, define $\mathcal{Y}^{+}=(Y_0, Y_1, Y_2, ...)$ and note that for $\mathcal{O}_{t}^{0}$ to occur it is necessary that $t>t_0$ and $\mathcal{Y}^{+}=(A,\ldots)$. Initially we list the events $C_k^{+}$ for $1 \leq k \leq 5$ and present in Fig. \ref{fig_deposition} a scheme illustrating the occurrence of those.
\begin{align*}
	&C_1^{+} = \{t>t_0, \ t_0<t_1\} \cap  \{\mathcal{Y}^{+}=(A, ...)\}, \\ 
	&C_2^{+} = \{t>t_0, \ t_0>t_1\} \cap \{\mathcal{Y}^{+}=(A, A, ...)\}, \\ 
	&C_3^{+} = \{t>t_0, \ t_0>t_1>t_2,\ t_2<t_3\} \cap \{\mathcal{Y}^{+}=(A, B, A, ...)\}, \\ 
	&C_4^{+} = \{t>t_0, \ t_0>t_1>t_2>t_3\} \cap \{\mathcal{Y}^{+}=(A, B, A, A, ...)\}, \\ 
	&C_5^{+} = \{t>t_0, \ t_0>t_1>t_2>t_3>t_4,\ t_4<t_5\} \cap \{\mathcal{Y}^{+}=(A, B, A, B, A, ...)\}.
\end{align*}

\noindent To define $C_k^+$ for any $k\geq0$, consider $\mathcal{H}_k^{+}=\{t>t_0>t_1>...>t_{k}\}$,
\begin{align*}
Z_{2k+1}=(\underbrace{\underline{A, B,} \ \underline{A, B,} \ ... \ , \underline{A, B,}}_{\mbox{\small{$k$ times}}} \ A, ...), \mbox{ and} \ \ Z_{2k+2}=(\underbrace{\underline{A, B,} \ \underline{A, B,} \ ... \ , \underline{A, B,}}_{\mbox{\small{$k$ times}}} \ A, A, ...).
\end{align*}
\noindent Then, 
        $$C_{2k+1}^{+} =  \mathcal{H}_{2k}^{+} \cap \{t_{2k}<t_{2k+1}\} \cap \{\mathcal{Y}^{+}=Z_{2k+1}\}$$
and 
	$$C_{2k+2}^{+} = \mathcal{H}_{2k+1}^{+} \cap \{\mathcal{Y}^{+}=Z_{2k+2}\}.$$

Analogously, considering the model restricted to sites $\{0,-1, -2,\ldots\}$ we can define similar events, which will be denoted by $C_{k}^{-}$, so that the occurrence of any one of them implies the occurrence of $\mathcal{O}_{t}^{0}$. Defining $\mathcal{Y}^{-}~=~(Y_{0}, Y_{-1}, Y_{-2},\ldots)$ and
\begin{align*}
	\mathcal{H}_k^{-}=\{t>t_0>t_{-1}>...>t_{-k}\},
\end{align*}
\noindent we have that, for $k\geq0$, 
	$$C_{2k+1}^{-} = \mathcal{H}_{2k}^{-} \cap \{t_{-2k}<t_{-2k-1}\} \cap \{\mathcal{Y}^{-}=Z_{2k+1}\}$$ 
 \noindent and
	$$C_{2k+2}^{-} = \mathcal{H}_{2k+1}^{-}\cap \{\mathcal{Y}^{-}=Z_{2k+2}\}.$$

Considering the model on $\mathbb{Z}$, observe that for $\mathcal{O}_{t}^{0}$ to occur it is necessary and sufficient that $C_{j}^{-} \cap C_{k }^{+}$ occurs for some $j, k \geq 1$. The definition of the events $C^{-}_{j}$ and $C^{+}_{k}$, as well as the calculation of their respective probabilities, vary depending on the parity of $j$ and $k$. Therefore, for each $j,k \geq 0$, we define the following events (main events) that will be used in the calculation of $\rho_{A}(t; \alpha)$:
\begin{itemize}
\item $\mathcal{G}_{j,k}^{ee}=C_{2j+2}^{-} \cap C_{2k+2}^{+}$;
\item $\mathcal{G}_{j,k}^{oe}=C_{2j+1}^{-} \cap C_{2k+2}^{+}$;
\item $\mathcal{G}_{j,k}^{eo}=C_{2j+2}^{-} \cap C_{2k+1}^{+}$;
\item $\mathcal{G}_{j,k}^{oo}=C_{2j+1}^{-} \cap C_{2k+1}^{+}$.
\end{itemize}

\noindent The superscripts $e$ (even) and $o$ (odd) in the notation of the main events refer to the parity of the subscript of the events $C^{-}_j$ and $C^{+}_k$, respectively. In this way, we can write
\begin{align}
	\mathcal{O}_{t}^{0}&=\bigcup_{j,k \geq 1} C_{j}^{-} \cap C_{k}^{+} \nonumber \\ 
	&=\left(\bigcup_{j,k \geq 0} \mathcal{G}_{j,k}^{ee}\right) \textstyle{\bigcup} \displaystyle\left(\bigcup_{j,k \geq 0} \mathcal{G}_{j,k}^{oe}\right) \textstyle{\bigcup} \left(\displaystyle\bigcup_{j,k \geq 0} \mathcal{G}_{j,k}^{eo}\right) \textstyle\bigcup \left(\displaystyle\bigcup_{j,k \geq 0} \mathcal{G}_{j,k}^{oo}\right),
\end{align}

\noindent where all events in the last row are mutually exclusive. Therefore, we have that
\begin{align}
	\rho_{A}(t; \alpha) = \sum_{j,k \geq 0} \left[ \mathbb{P}(\mathcal{G}_{j,k}^{ee}) + \mathbb{P}(\mathcal{G}_{j,k}^{oe}) + \mathbb{P}(\mathcal{G}_{j,k}^{eo}) + \mathbb{P}(\mathcal{G}_{j,k}^{oo}) \right].
	\label{somaprob}
\end{align}

\vskip 1cm

\section{Probability of the main events}

We are now going to determine the value of each probability of the expression (\ref{somaprob}). Note that calculating the probability of $\mathcal{Y}^{+}=Z_{k}$ (or $\mathcal{Y}^{-}=Z_{k}$) is straightforward; if $n_A$ and $n_B$ represent the amount of $A$'s and $B$'s we fixed, then this probability is $\alpha^{n_A}\beta^{n_B}$. In general, we have that
\begin{align*}
	\mathbb{P}(\mathcal{Y}^{-}=Z_{2j+2}, \ \mathcal{Y}^{+}=Z_{2k+2})=(\alpha^{j+2}\beta^{j})\cdot (\alpha^{k+2}\beta^{k})\cdot \alpha^{-1} = \alpha^{j+k+3}\beta^{j+k}, \\ 
	\mathbb{P}(\mathcal{Y}^{-}=Z_{2j+1}, \ \mathcal{Y}^{+}=Z_{2k+2})=(\alpha^{j+1}\beta^{j})\cdot (\alpha^{k+2}\beta^{k})\cdot \alpha^{-1} = \alpha^{j+k+2}\beta^{j+k}, \\
	\mathbb{P}(\mathcal{Y}^{-}=Z_{2j+2}, \ \mathcal{Y}^{+}=Z_{2k+1})=(\alpha^{j+2}\beta^{j})\cdot (\alpha^{k+1}\beta^{k})\cdot \alpha^{-1} = \alpha^{j+k+2}\beta^{j+k},
\end{align*}
\noindent and
\begin{align*}
	\mathbb{P}(\mathcal{Y}^{-}=Z_{2j+1}, \ \mathcal{Y}^{+}=Z_{2k+1})=(\alpha^{j+1}\beta^{j})\cdot (\alpha^{k+1}\beta^{k})\cdot \alpha^{-1} = \alpha^{j+k+1}\beta^{j+k}.
\end{align*}

\noindent For the events related to the variable $t$, the idea is to decompose them as $\mathcal{H}^{-}_{j} \cap \mathcal{H}^{+}_{k}$ and use the fact that
\begin{align}	
\mathbb{P}\left(\mathcal{H}_{j}^{-},\ \mathcal{H}_{k}^{+}\right)&=
 \int_0^t \mathbb{P}\left(u>t_{-1}>\ldots>t_{-j},\ u>t_1>\ldots>t_{k}\right) du\nonumber \\
 &=\int_0^t \mathbb{P}\left(u>t_{-1}>\ldots>t_{-j}\right)\mathbb{P}\left( u>t_1>\ldots>t_{k}\right) du\nonumber\\
&=\int_0^t \frac{u^{j}}{j!}\frac{u^{k}}{k!} \ du=\frac{t^{j+k+1}}{j!k!(j+k+1)}.\label{probabHH}
\end{align}

\noindent We will sometimes use that for any two events $\mathcal{B}$ and $\mathcal{D}$, it is hold that
\begin{align}\label{fundamental}
	\mathbb{P}(\mathcal{B}\cap\mathcal{D})=\mathbb{P}(\mathcal{B})-\mathbb{P}(\mathcal{B}\cap\mathcal{D}^c).
\end{align}

\noindent Furthermore, consider $\alpha \neq 0$ and $\alpha \neq 1$ and note that it is simple to deduce that $\rho_{A}(t;0)=0$ and $\rho_{A}(t;1)=t$ for all $t \in [0,1]$. Next, we calculate the value of each of the probabilities in~(\ref{somaprob}).

\vskip 0.7cm

\subsection{Probability $\mathbb{P}(\mathcal{G}_{j,k}^{ee})$ }

We have that
\begin{align*}
	\mathbb{P}(\mathcal{G}_{j,k}^{ee})&= \mathbb{P}(C^{-}_{2j+2}, \ C^{+}_{2k+2}) =\mathbb{P}(\mathcal{Y}^{-}=Z_{2j+2}, \ \mathcal{Y}^{+}=Z_{2k+2}, \ \mathcal{H}_{2j+1}^{-}, \ \mathcal{H}_{2k+1}^{+}) \\
        &=\mathbb{P}(\mathcal{Y}^{-}=Z_{2j+2}, \ \mathcal{Y}^{+}=Z_{2k+2}) \cdot \mathbb{P}(\mathcal{H}_{2j+1}^{-}, \ \mathcal{H}_{2k+1}^{+})
\end{align*}

\noindent Therefore,
\begin{align*}
	\dfrac{1}{\alpha^{j+k+3}\beta^{j+k}} \cdot \mathbb{P}(\mathcal{G}_{j,k}^{ee})&=\mathbb{P}(\mathcal{H}_{2j+1}^{-}, \ \mathcal{H}_{2k+1}^{+})
\end{align*}

\noindent and from Eq. \eqref{probabHH}, 
\begin{align*}
	\dfrac{1}{\alpha^{j+k+3}\beta^{j+k}} \cdot \mathbb{P}(\mathcal{G}_{j,k}^{ee})& = \frac{t^{2j+2k+3}}{(2j+1)!(2k+1)!(2j+2k+3)}.
\end{align*}

\noindent To shorten our expressions, we denote $\gamma=\sqrt{\alpha \beta}$ and
$$\Gamma_{j,k}(m,n)=\dfrac{(\gamma t)^{2j+2k+m+n+1}}{(2j+m)!(2k+n)!(2j+2k+m+n+1)}.$$ 

\noindent In this way, we can write
\begin{align*}
	\mathbb{P}(\mathcal{G}_{j,k}^{ee})&= \left(\dfrac{\alpha}{\beta}\right)^{\frac{3}{2}} \cdot \frac{(\gamma t)^{2j+2k+3}}{(2j+1)!(2k+1)!(2j+2k+3)} = \left(\dfrac{\alpha}{\beta}\right)^{\frac{3}{2}} \cdot \Gamma_{j,k}(1,1).
\end{align*}

\vskip 0.7cm

\subsection{Probability $\mathbb{P}(\mathcal{G}_{j,k}^{oe})$} 

According to the definition of $\mathcal{G}_{j,k}^{oe}$, we have that
\begin{align*}
	\mathbb{P}(\mathcal{G}_{j,k}^{oe})=\mathbb{P}(\mathcal{Y}^{-}=Z_{2j+1}, \ \mathcal{Y}^{+}=Z_{2k+2}, \ \mathcal{H}_{2j}^{-}, \ t_{-2j}<t_{-2j-1}, \ \mathcal{H}_{2k+1}^{+})
\end{align*}

\noindent Therefore,
\begin{align*}
	\dfrac{1}{\alpha^{j+k+2}\beta^{j+k}} \cdot \mathbb{P}(\mathcal{G}_{j,k}^{oe})&=\mathbb{P}(\mathcal{H}_{2j}^{-}, \ t_{-2j}<t_{-2j-1}, \ \mathcal{H}_{2k+1}^{+}). 
\end{align*}

\noindent Through Eq. \eqref{fundamental}, we can write
\begin{align*}
	\mathbb{P}(\mathcal{H}_{2j}^{-}, \ t_{-2j}<t_{-2j-1}, \ \mathcal{H}_{2k+1}^{+}) &=\mathbb{P}(\mathcal{H}_{2j}^{-}, \ \mathcal{H}_{2k+1}^{+})-\mathbb{P}(\mathcal{H}_{2j}^{-}, \ \mathcal{H}_{2k+1}^{+}, \ t_{-2j}>t_{-2j-1}) \\
	&=\mathbb{P}(\mathcal{H}_{2j}^{-}, \ \mathcal{H}_{2k+1}^{+})-\mathbb{P}(\mathcal{H}_{2j+1}^{-}, \ \mathcal{H}_{2k+1}^{+}).
\end{align*}
Using the previous equation and the Eq. \eqref{probabHH}, we derive that
\begin{align*}
	\dfrac{1}{\alpha^{j+k+2}\beta^{j+k}} \cdot \mathbb{P}(\mathcal{G}_{j,k}^{oe})=\frac{t^{2j+2k+2}}{(2j)!(2k+1)!(2j+2k+2)}-\frac{t^{2j+2k+3}}{(2j+1)!(2k+1)!(2j+2k+3)}.
\end{align*}
After some algebraic manipulations, we finally obtain
\begin{align*}
	\mathbb{P}(\mathcal{G}_{j,k}^{oe})&= \left(\dfrac{\alpha}{\beta}\right) \frac{(\gamma t)^{2j+2k+2}}{(2j)!(2k+1)!(2j+2k+2)}-\left(\dfrac{\alpha^{\frac{1}{2}}}{\beta^{\frac{3}{2}}} \right) \frac{(\gamma t)^{2j+2k+3}}{(2j+1)!(2k+1)!(2j+2k+3)} \\
        &=\dfrac{\alpha}{\beta} \cdot \Gamma_{j,k}(0,1) - \dfrac{\alpha^{\frac{1}{2}}}{\beta^{\frac{3}{2}}} \cdot \Gamma_{j,k}(1,1).
\end{align*}

\vskip 0.7cm

\subsection{Probability $\mathbb{P}(\mathcal{G}_{j,k}^{eo})$}

This calculation follows the same steps as those performed in the previous subsection. One can use Eqs. \eqref{probabHH} and \eqref{fundamental} to verify that
\begin{align*}
	\dfrac{1}{\alpha^{j+k+2}\beta^{j+k}} \cdot \mathbb{P}(\mathcal{G}_{j,k}^{eo})&=\mathbb{P}(\mathcal{H}_{2j+1}^{-}, \ \mathcal{H}_{2k}^{+}, \ t_{2k}<t_{2k+1}) \\
	&=\mathbb{P}(\mathcal{H}_{2j+1}^{-}, \ \mathcal{H}_{2k}^{+})-\mathbb{P}(\mathcal{H}_{2j+1}^{-}, \ \mathcal{H}_{2k}^{+}, \ t_{2k}>t_{2k+1}) \\
	&=\mathbb{P}(\mathcal{H}_{2j+1}^{-}, \ \mathcal{H}_{2k}^{+})-\mathbb{P}(\mathcal{H}_{2j+1}^{-}, \ \mathcal{H}_{2k+1}^{+}) \\
	&=\frac{t^{2j+2k+2}}{(2j+1)!(2k)!(2j+2k+2)}-\frac{t^{2j+2k+3}}{(2j+1)!(2k+1)!(2j+2k+3)}.
\end{align*}
Therefore,
\begin{align*}
	\mathbb{P}(\mathcal{G}_{j,k}^{eo})&= \dfrac{\alpha}{\beta} \cdot \frac{(\gamma t)^{2j+2k+2}}{(2j+1)!(2k)!(2j+2k+2)}- \dfrac{\alpha^{\frac{1}{2}}}{\beta^{\frac{3}{2}}} \cdot \frac{(\gamma t)^{2j+2k+3}}{(2j+1)!(2k+1)!(2j+2k+3)} \\ 
	&=\dfrac{\alpha}{\beta} \cdot \Gamma_{j,k}(1,0) - \dfrac{\alpha^{\frac{1}{2}}}{\beta^{\frac{3}{2}}} \cdot \Gamma_{j,k}(1,1).
\end{align*}

\vskip 0.7cm

\subsection{Probability $\mathbb{P}(\mathcal{G}_{j,k}^{oo})$}

Note that 
\begin{equation*}
\dfrac{1}{\alpha^{j+k+1}\beta^{j+k}} \cdot \mathbb{P}(\mathcal{G}_{j,k}^{oo})=\mathbb{P}(\mathcal{H}_{2j}^{-}, \ t_{-2j}<t_{-2j-1}, \ \mathcal{H}_{2k}^{+}, \ t_{2k}<t_{2k+1}).    
\end{equation*}
Next, using Equation \eqref{fundamental} twice,  we obtain that
    \begin{align*}
	\dfrac{1}{\alpha^{j+k+1}\beta^{j+k}} \cdot \mathbb{P}(\mathcal{G}_{j,k}^{oo})
	&=\mathbb{P}(\mathcal{H}_{2j}^{-}, \ t_{-2j}<t_{-2j-1}, \ \mathcal{H}_{2k}^{+}) - \mathbb{P}(\mathcal{H}_{2j}^{-}, \ t_{-2j}<t_{-2j-1}, \ \mathcal{H}_{2k+1}^{+}) \\
	&=[\mathbb{P}(\mathcal{H}_{2j}^{-}, \ \mathcal{H}_{2k}^{+}) - \mathbb{P}(\mathcal{H}_{2j+1}^{-}, \ H_{2k}^{+})]-[\mathbb{P}(\mathcal{H}_{2j}^{-}, \ H_{2k+1}^{+}) - \mathbb{P}(\mathcal{H}_{2j+1}^{-}, \ \mathcal{H}_{2k+1}^{+})].
\end{align*}
Hence, Eq. \eqref{probabHH} gives us that
\begin{align*}
	\dfrac{1}{\alpha^{j+k+1}\beta^{j+k}} \cdot \mathbb{P}(\mathcal{G}_{j,k}^{oo})&=
	\frac{t^{2j+2k+1}}{(2j)!(2k)!(2j+2k+1)}-\frac{t^{2j+2k+2}}{(2j+1)!(2k)!(2j+2k+2)} \\
	&-\frac{t^{2j+2k+2}}{(2j)!(2k+1)!(2j+2k+2)}+\frac{t^{2j+2k+3}}{(2j+1)!(2k+1)!(2j+2k+3)}.
\end{align*}
Therefore, 
\begin{align*}
	\mathbb{P}(\mathcal{G}_{j,k}^{oo})&= \left(\dfrac{\alpha}{\beta}\right)^{\frac{1}{2}} \cdot \frac{(\gamma t)^{2j+2k+1}}{(2j)!(2k)!(2j+2k+1)} - \dfrac{1}{\beta} \cdot \frac{(\gamma t)^{2j+2k+2}}{(2j+1)!(2k)!(2j+2k+2)} \\
	&-\dfrac{1}{\beta} \cdot \frac{(\gamma t)^{2j+2k+2}}{(2j)!(2k+1)!(2j+2k+2)} +\dfrac{1}{\alpha^{\frac{1}{2}} \beta^{\frac{3}{2}}} \cdot \frac{(\gamma t)^{2j+2k+3}}{(2j+1)!(2k+1)!(2j+2k+3)} \\ 
	&= \left(\dfrac{\alpha}{\beta}\right)^{\frac{1}{2}} \cdot \Gamma_{j,k}(0,0) - \dfrac{1}{\beta} \cdot \Gamma_{j,k}(1,0) - \dfrac{1}{\beta} \cdot \Gamma_{j,k}(0,1) +\dfrac{1}{\alpha^{\frac{1}{2}} \beta^{\frac{3}{2}}} \cdot \Gamma_{j,k}(1,1).
\end{align*}

\section{Calculation of $\rho_{A}(t;\alpha)$}

First, remember that
\begin{align}
	\rho_{A}(t; \alpha) = \sum_{j,k \geq 0} \left[ \mathbb{P}(\mathcal{G}_{j,k}^{ee}) + \mathbb{P}(\mathcal{G}_{j,k}^{oe}) + \mathbb{P}(\mathcal{G}_{j,k}^{eo}) + \mathbb{P}(\mathcal{G}_{j,k}^{oo}) \right].
	\label{somaprob}
\end{align}
Next, substituting the expressions obtained for the main events probabilities and $\alpha=1-\beta$, we derive that
\begin{align}
	\begin{split}
&\mathbb{P}(\mathcal{G}_{j,k}^{ee}) + \mathbb{P}(\mathcal{G}_{j,k}^{oe}) + \mathbb{P}(\mathcal{G}_{j,k}^{eo}) + \mathbb{P}(\mathcal{G}_{j,k}^{oo}) = \\
& =\left(\dfrac{\alpha}{1-\alpha}\right)^{\frac{1}{2}} \cdot \Gamma_{j,k}(0,0) - \Gamma_{j,k}(1,0) - \Gamma_{j,k}(0,1) + \left(\dfrac{1-\alpha}{\alpha}\right)^{\frac{1}{2}} \cdot \Gamma_{j,k}(1,1).
	\end{split}
\label{sum_prob}
\end{align}

\noindent The summation in $j,k$ of $\Gamma_{j,k}(m,n)$, for each $m,n \in \{0,1\}$, is then determined by expanding the hyperbolic sines and cosines in power series. Indeed,
\begin{align}
\begin{split}
\sum_{j, k \geq 0} \Gamma_{j,k}(0,0)
&=\sum_{j,k\geq0}\dfrac{(\gamma t)^{2k+2j+1}}{(2k)!(2j)!(2k+2j+1)} =
\sum_{k \geq 0}\sum_{j\geq0} \int_0^{\gamma t}\dfrac{u^{2k+2j}}{(2k)!(2j)!}\ du \\
&=\sum_{k \geq 0} \int_0^{\gamma t}\dfrac{u^{2k}}{(2k)!}\sum_{j\geq0}\dfrac{u^{2j}}{(2j)!}\ du =\sum_{k \geq 0} \int_0^{\gamma t}\dfrac{u^{2k}}{(2k)!}\cosh u\ du \\ 
&=\int_0^{\gamma t} \cosh^2 u\ du. 
\label{sum_gamma1}
\end{split}
\end{align}
In a similar manner, it can be verified that 
\begin{align}
	\begin{split}
		\sum_{j, k \geq 0} \Gamma_{j,k}(0,1)=\sum_{k \geq 0}\int_0^{\gamma t} \dfrac{u^{2k+1}}{(2k+1)!}\cosh{u} \ du = \int_0^{\gamma t} \sinh(u)\cosh{(u)} \ du; 
	\end{split}
	\label{sum_gamma2}
\end{align}
\begin{align}
	\begin{split}
		\sum_{j, k \geq 0} \Gamma_{j,k}(1,0)=\sum_{k \geq 0}\int_0^{\gamma t}\dfrac{u^{2k}}{(2k)!}\sinh{u} \ du 
		= \int_0^{\gamma t} \cosh(u)\sinh{(u)} \ du; 
	\end{split}
	\label{sum_gamma3}
\end{align}
\noindent and
\begin{align}
	\begin{split}
		\sum_{j, k \geq 0} \Gamma_{j,k}(1,1)=\sum_{k \geq 0}\int_0^{\gamma t}\dfrac{u^{2k+1}}{(2k+1)!}\sinh{u} \ du = \int_0^{\gamma t} \sinh^2{u}.
	\end{split}
	\label{sum_gamma4}
\end{align}

\noindent Therefore, defining $\theta=\sqrt{\dfrac{\alpha}{1-\alpha}}$, from Eqs. (\ref{somaprob}) to (\ref{sum_gamma4}) we obtain that 

\begin{figure}[t!]
	\begin{center}
		\includegraphics[width=7.9cm, height=7.9cm]{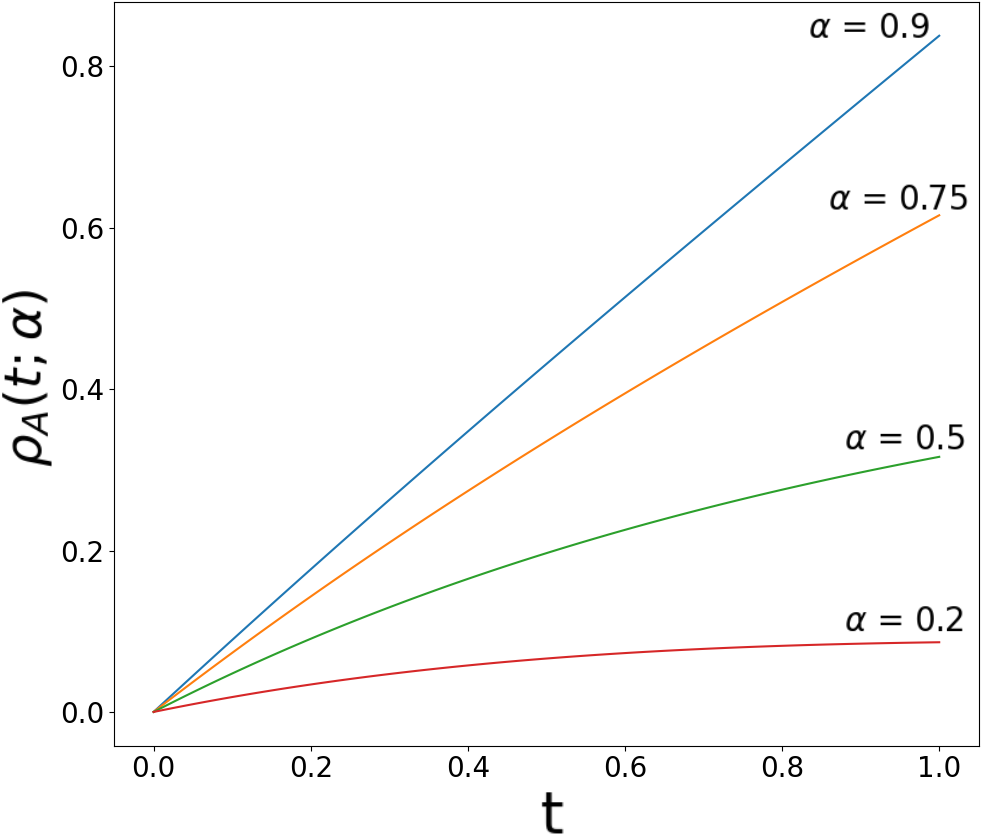} \ \
        \includegraphics[width=7.9cm, height=7.9cm]{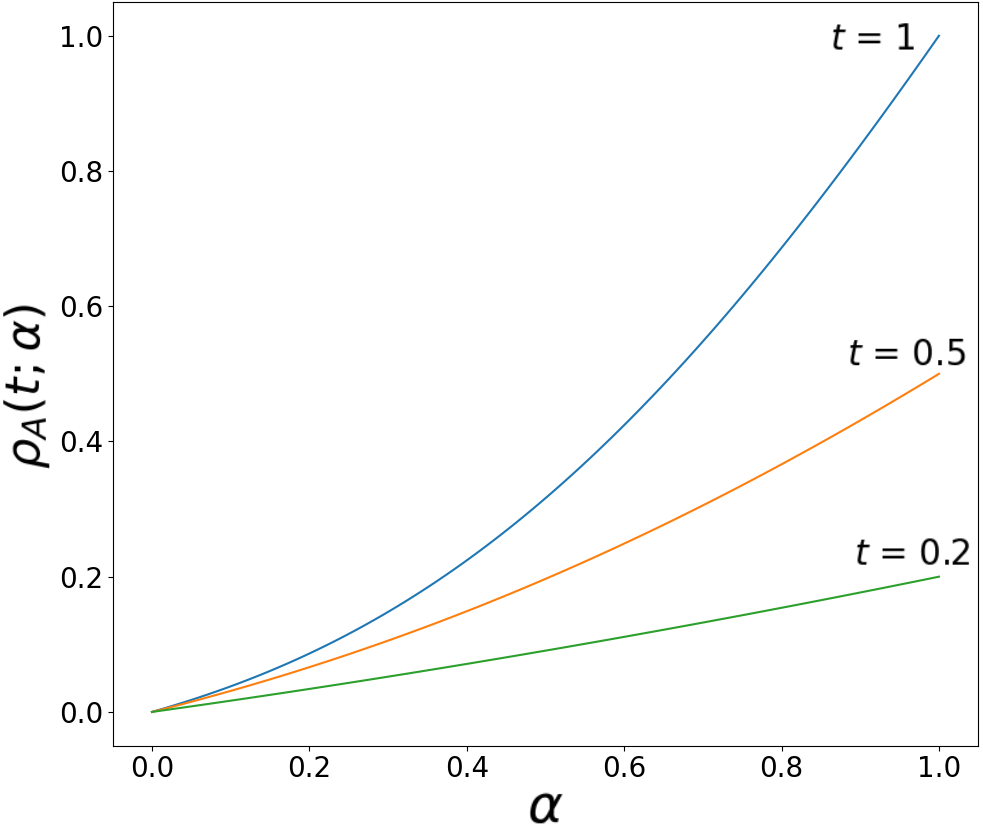}
		\caption{Graph of average densities $\rho_{A}(t;\alpha)$ for $\alpha=0.2$, $\alpha=0.5$, $\alpha=0.75$ and $\alpha=0.9$ (left panel); and for $t=0.2$, $t=0.5$ and $t=1$ (right panel).}
		\label{fig_density}
	\end{center}
\end{figure}

\begin{figure}[t!]
	\begin{center}
		\includegraphics[width=11cm, height=10cm]{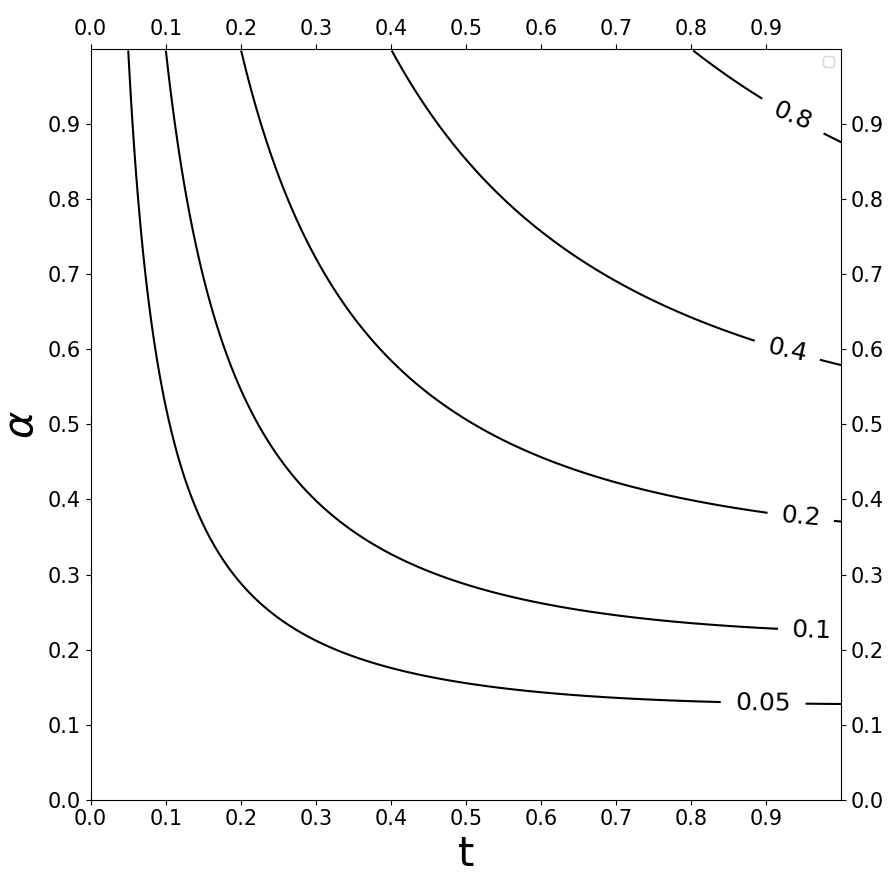}
		\caption{Contour curves of $\rho_{A}(t;\alpha)=\lambda$ for $\lambda \in \{0.05, \ 0.1, \ 0.2, \ 0.4, \ 0.8\}$. The values of $\lambda$ corresponding to each curve are indicated in the figure.}
		\label{c_n}
	\end{center}
\end{figure}

\begin{align*}
	\rho_{A}(t; \alpha) &= \sum_{j,k \geq 0} \left[ \mathbb{P}(\mathcal{G}_{j,k}^{oo}) + \mathbb{P}(\mathcal{G}_{j,k}^{eo}) + \mathbb{P}(\mathcal{G}_{j,k}^{oe}) + \mathbb{P}(\mathcal{G}_{j,k}^{ee}) \right] \\
	& = \sum_{j,k \geq 0} \left[ \theta \cdot \Gamma_{j,k}(0,0) - \Gamma_{j,k}(1,0) - \Gamma_{j,k}(0,1) + \dfrac{1}{\theta} \cdot \Gamma_{j,k}(1,1) \right] \\
	& = \int_0^{\gamma t}\left( \theta \cosh^2 u -2 \sinh u \cosh u + \dfrac{1}{\theta} \sinh^2 u \right) du\\
 & = \int_0^{\gamma t}\left( \sqrt{\theta} \cosh u -  \dfrac{1}{\sqrt{\theta}} \sinh u \right)^2 du.
\end{align*}
Finally, by solving the above integral, we get the desired,
\begin{align*}
	\rho_{A}(t; \alpha) =  \dfrac{1}{4\theta} \left[2\theta+ \left(\theta^2 - 1\right)2\gamma t + \left(\theta^2+1\right)\sinh(2\gamma t) - 2\theta\cosh(2\gamma t) \right].
\end{align*}

\section{Behavior of $\rho_{A}(t; \alpha)$}

In Fig. \ref{fig_density} (left panel) we present the graph of $\rho_{A}(t;\alpha)$ for some values of $\alpha$. Note that for $\alpha=0.9$ ($\beta=0.1$) the growth of the average density of sites $A$ is almost linear, this occurs because only a few particles $B$ are deposited during the process ($\rho_ {B}(1;0.9) \approx 0.038$). Thus, most attempts to depose particles of the $A$ type are successful ($\rho_{A}(1;0.9) \approx 0.84$). Similar reasoning can be applied to explain the average density behavior observed for low $\alpha$ values (e.g., $\alpha=0.2$). As the process progresses, the rate of particle deposition for type $A$ diminishes due to the increasing number of depositions for particles of type $B$. This interference from particle $B$ depositions limits the number of successful depositions for particles of type $A$. Consequently, as $t$ approaches $1$, the average density $\rho_{A}$ reaches a near-constant value. 

In the right panel of Fig. \ref{fig_density} the average density value is presented as a function of $\alpha$ for three different values of $t$. For low $t$, such as $t=0.2$, we observe that $\rho_{A}(t;\alpha) \approx  \alpha t$; as with only a few particles deposited, almost all deposition attempts are successful. This behavior becomes evident when we examine the first terms of the series expansion of the function $\rho_{A}(t;\alpha)$ about $t=0$:
\begin{align}
    \rho_{A}(t;\alpha)=\alpha t + \alpha(\alpha-1) t^2 +\dfrac{\alpha(1-\alpha)}{3} t^3 - \dfrac{\alpha^2(1-\alpha)^2}{3} t^4 + \mathcal{O}(t^5).
\end{align}

When $t=1$, the behavior of the average density for small $\alpha$ is given by the series expansion of $\rho_{A}(1;\alpha)$ about $\alpha=0$:
\begin{align}
    \rho_{A}(1;\alpha) =  \dfrac{1}{3}\alpha + \dfrac{2}{5} \alpha^2 + \dfrac{52}{105} \alpha^3 - \dfrac{88}{567} \alpha^4 + \mathcal{O}(\alpha^{\frac{9}{2}}).
\end{align}

The Fig. \ref{c_n} displays the contour curves $\rho_{A}(t;\alpha)=\lambda$ for various values of $\lambda$. When $\alpha=1$, we have $t=\lambda$. However, when $t=1$, we observe that $\alpha>\lambda$ for $\alpha \in (0,1)$, as the deposition of particles of type $B$ hinders some attempts of deposition of particles $A$. This relationship between $\alpha$ and $\lambda$ at $t=1$ can be observed in the right panel of Fig. \ref{fig_density}.

\section{Conclusion}

In this paper, we have investigated the \textit{One-dimensional $AB$ random sequential adsorption with one deposition per site} in which particles of types $A$ and $B$ are sequentially added to a lattice with the constraint that opposite types cannot occupy nearest-neighbor sites. Considering a single deposition attempt per site, we have analytically calculated the average density of sites occupied by particles of type $A$, $B$ and empty sites at all time instances, and for all deposition probabilities of each particle type.

Our analytical calculations provide valuable insights into the average density dynamics of particles $A$ in the studied model, shedding light on the interplay between deposition attempts, particle types, and time evolution. These findings contribute to the broader field of random sequential adsorption and may inspire further investigations in other $AB$ RSA models. Future research can focus on studying this model in higher dimensions and investigating the effects of different deposition rules on the system's behavior.

\section*{Acknowledgments}

We thank Robert Ziff for his helpful suggestions and comments to improve the manuscript. D.C.S. was partially supported by CNPq (Grant 409198/2021-8) and FAPEMIG (Processo APQ-00774-21).


\begin{thebibliography}{00}
		\bibliographystyle{iopart-num}

		\bibitem{ramsden_thin_film1}
		J. J. Ramsden, \emph{J. Stat. Phys.} 73 (1993) 853-877.
		
		\bibitem{sorrell_thin_film2}
		C. D. Sorrell, L. A. Lyon,  \emph{Langmuir} 24(14) (2008) 7216-7222.
		
		\bibitem{krugmann_membrane1}
		B. Krugmann, A. Koutsioubas, L. Haris, S. Micciulla, D. Lairez, A. Radulescu, S. Förster and A. M. Stadler, \emph{Front. Chem.} 9 (2021) 631277.
		
		\bibitem{Lu_membrane2}
		R. Lu, Q. Li, T. H. Nguyen, \emph{J. Colloid Interface Sci.} 466 (2016) 120-127.
		
		\bibitem{carton_membrane3}
		I. Carton, A. R. Brisson, R. P. Richter, \emph{Anal. Chem.} 82(22) (2010) 9275-9281.
		
		
		\bibitem{molecules1_Li}
		J. Li, I. Tezsevin, M. J. M. Merkx, J. F. W. Maas, W. M. M. Kessels, T. E. Sandoval, A. J. M. Marckus, \emph{J. Vac. Sci. Technol.} 40(6) (2022) 062409.
		
		\bibitem{molecules2_Erban}
		R. Erban, S. J. Chapman, \emph{Phys. Rev. E} 75(4) (2007) 041116.
		
		\bibitem{molecules3_ciesla1}
		M. Cieśla, J. Barbasz, \emph{J. Mol. Model.} 19 (2013). 5423-5427.
		
		\bibitem{molecules4_ciesla2}
		M. Cieśla, G. Paj\c{a}k, R. M. Ziff, \emph{Phys. Chem. Chem. Phys.} 17(37) (2015). 24376-24381.
		
		\bibitem{penrose}
		M. Penrose, A. W. Sudbury, \emph{Ann. Appl. Probab.} 15(1B) (2005) 853–889.
		
		\bibitem{evans}
		J. W. Evans, \emph{Rev. Mod. Phys.} 130(5) (1993) 1281–1327.
		
		\bibitem{martins}
		P. H. Martins, R. Dickman, R. M. Ziff,  \emph{Phys. Rev. E}, 107(2) (2023) 024104.

        \bibitem{gerin}
        L. Gerin, \emph{Electron. J. Comb.}, 22(4), (2015).
  
		\bibitem{amaral_santos}
		C. S. do Amaral, D. C. dos Santos, Density and correlation in a random sequential adsorption model, arXiv preprint arXiv:2210.05627 (2022).
		
				
	\end{thebibliography}
\end{document}